\documentclass[%
 reprint,
superscriptaddress,
 amsmath,amssymb,
 aps,
pra,
]{revtex4-1}
\usepackage{graphicx,hyperref}
\usepackage{dcolumn}
\usepackage{bm}

\usepackage{lineno}

\newcommand{\lbl}{\label}
\newcommand{\bn}{\begin{align}}
\newcommand{\enl}[1]{\label{#1}\end{align}}

\newcommand{\ba}{\begin{equation}}
\newcommand{\bs}{\begin{split}}
\newcommand{\ea}{\end{equation}}
\newcommand{\el}[1]{\label{#1}\end{equation}}
\newcommand{\bsa}{\begin{subequations}}
\newcommand{\esa}{\end{subequations}}
\newcommand{\esl}[1]{\label{#1}\end{subequations}}
\newcommand{\baq}{\begin{eqnarray}}
\newcommand{\eaq}{\end{eqnarray}}

\newcommand{\eq}[1]{Eq.~(\ref{#1})}

\newcommand{\bra}{\langle}
\newcommand{\ket}{\rangle}

\newcommand{\+}{\dagger}

\newcommand{\dv}{\mathrm{d}}

\newcommand{\pd}[2]{\frac{\partial #1}{\partial #2}}

\newcommand{\um}[2]{\sum\limits_{#1}^{#2}}

\newcommand{\g}{\gamma}

\newcommand{\de}{\delta}
\newcommand{\De}{\Delta}
\newcommand{\E}{\epsilon}

\newcommand{\w}{\omega}
\newcommand{\W}{\Omega}

\begin{document}
\title{Light as quantum back-action nullifying meter}
\author{Sankar Davuluri}\email{sankar@hyderabad.bits-pilani.ac.in}\affiliation{Department of Physics, Birla Institute of Technology and Science, Pilani, Hyderabad Campus, Hyderabad 500078, India.}
\author{Yong Li}
\affiliation{Beijing Computational Science Research Center, Beijing 100193, China.}
\affiliation{Synergetic Innovation Center for Quantum Effects and Applications, Hunan Normal University, Changsha 410081, China.}
\date{\today}
\begin{abstract}
We propose a new method to overcome quantum back-action in a measurement process using oscillators. An optical oscillator is used as a meter to measure the parameters of another open oscillator. The optical oscillator is synthesized such that the optical restoring force counters any perturbations induced by the quantum back-action phenomena. As a result, it is shown that the quantum back-action in continuous measurement is suppressed in the low frequency regime i.e., for frequencies much smaller than the resonance frequency of the open oscillator. As the meter plays the role of measuring parameters as well as suppressing the quantum back-action, we call it as quantum back-action nullifying meter. As an application of this method, synthesis of quantum back-action nullifying optical oscillator for suppressing radiation pressure force noise in linear and non-linear optomechanics is described.
\end{abstract}

\maketitle

\section{Introduction}
Classical mechanics tells us that measurements can be infinitely accurate given the measurement scheme and the apparatus are perfect. On contrary to that, quantum mechanics imposes limits on the accuracy of our measurements via the Heisenberg uncertainty principle. The uncertainty principle~\cite{clerk-09,zettili-text}, which gives the fundamental quantum limit, is neither a consequence of the measuring device nor the measurement strategy. While the uncertainty principle is unbeatable, other quantum limits such as standard quantum limit are avoidable~\cite{braginsky-80,maccone}. Standard quantum limit is a consequence of quantum back action (QBA)~\cite{corbitt-19,schreppler,kippenberg-ba,caves-rmp,boyd-20} which states that an accurate measurement at a particular time induces uncertainty in the same measurement performed at a later time. In a continuous measurement, QBA limits the accuracy of our measurements atleast to standard quantum limit. Recent experimental advances have shown that QBA limits the measurement accuracy in many physical systems~\cite{schreppler,spethmann,corbitt-19,kim,ando}. Overcoming it requires performing a special kind of measurement known as quantum non-demolition measurement~\cite{unruh,braginsky-80}. However, a quantum non-demolition measurement is not possible in all the experiments as it requires the measured variable to commute with the Hamiltonian. Thus several alternative mechanisms~\cite{yuen,zubova,dowling,caves-10,schwab-10,kimble,sankar-ol,shomroni,lett,mason,yu2020} are being developed to overcome QBA. This article proposes a new method to suppress QBA by using a meter with intrinsic restoring force.
\par Quantum back-action in continuous measurement stems from the interplay between the canonically conjugate variables. Such an interplay can arise~\cite{edelstein} from the natural evolution of the system or because of the interaction between system and meter. Even though the physics of QBA can be illustrated~\cite{caves-int,bondurant} by using isolated systems like a quantum free particle or a quantum simple harmonic oscillator, these systems are never isolated in experiments. A system on which a measurement is to be performed is generally coupled to (i) environment and (ii) meter. In my previous work~\cite{sankar-ol}, decoherence from the environment was used to suppress the QBA by erasing the memory of previous measurement from the system. In this article, we propose a new method to suppress QBA by using the restoring force of the meter to counter the perturbation induced by the QBA.
\par In a measurement process, generally, the experimenter has control over the meter and its coupling to the system on which the measurement is to be performed. Evasion of QBA can be achieved by choosing a meter with intrinsic restoring force, an oscillator for example, and then using the restoring force to counter any perturbation induced by the QBA phenomena. As the meter in this method serves the dual purpose of measuring, and nullifying the QBA, we call it as quantum back-action nullifying meter (QBNM).
\par The theory of QBNM is developed by modeling the system and the meter as oscillators. Even though the meter could be any measuring device with intrinsic restoring force, we restrict the meter to an optical oscillator as optics offer unique advantage in the experimental implementation of this technique. As many systems~\cite{susanne,gil-santos,esslinger,shkarin} can be modeled with simple harmonic oscillators, the results presented in this article will be useful in many areas of physics.
\par Consider a simple harmonic oscillator (SHO) with momentum $\hat p_1$, position $\hat z_1$, mass $m_1$, and Eigen frequency $\w_1$.
The equations of motion are given as
\ba\dot{\hat p}_1=-m_1\w_1^2\hat z_1,\quad\quad m_1\dot{\hat z}_1=\hat p_1.\el{136}
Solving \eq{136} gives the position of the simple harmonic oscillator at time $t$ as $\hat z_1=\hat z_2\cos(\w_1t_e)+\hat p_2/(m_1\w_1)\sin(\w_1t_e),$ where $\hat z_2$ is the initial position, $\hat p_2$ is the initial momentum, and $t_e=t-t_2$ with $t_2$ as initial time. The $\hat z_2$ and $\hat p_2$ leads to initial imprecision (IIP) and QBA, respectively. 
The competition between IIP and QBA limits the accuracy in continuous measurement of $\hat z_1$ atleast to $\sqrt{\hbar/m_1\w_1}$, where $\hbar$ is the reduced Planck constant, which is the standard quantum limit. The QBA in \eq{136} arises from the canonically conjugate variables. We haven't considered any specific measurement scheme or measuring device in \eq{136}, but it already implies standard quantum limit. The simple harmonic oscillator considered until now is an isolated case. In an experiment, the simple harmonic oscillator is coupled to the environment, and a measurement can be performed by coupling it to a meter. 
In order to account for every thing in a measurement process, we have to include both the reservoir and the meter.
\par For simplicity, the simple harmonic oscillator coupled to reservoir will be called as an open simple harmonic oscillator (OSHO). Suppose that we are interested in measuring the position $\hat z_s$ of an OSHO whose momentum, mass, and Hamiltonian are represented by $\hat p_s$, $m_s$, and $\hat H_s$, respectively. Experimentally, $\hat z_s$ is estimated by coupling the OSHO to a meter. We assume that the meter is an optical oscillator with Hamiltonian $\hat H_o$, effective momentum $\hat p_o$, effective position $\hat z_o$ and effective mass $m_o$. The optical oscillator can be synthesized by driving an optical cavity with an external laser field. The optical field inside the cavity oscillates at the Eigen frequency $\w_o$ of the cavity and this creates an optical oscillator. As the optical oscillator is driven at frequency $\w_d$ of the driving laser, after making a unitary transformation with $\hat U=e^{i\w_d\hat H_ot/\hbar\w_o}$ the total Hamiltonian becomes $\hat H_s+\De\hat H_o/\w_o+\hat H_{os}+\hat H_r$, where $\De=\w_o-\w_d$, $\hat H_{os}$ is the Hamiltonian for interaction between optical oscillator and OSHO, and $\hat H_r$ is the Hamiltonian of the reservoirs and their interaction with the OSHO and the optical oscillator. The explicit form of the $\hat H_{os}$ depends on the nature of coupling between the OSHO and the meter. We keep the $\hat H_{os}$ arbitrary. The equations of motion are given as
\ba\dot{\hat z}_{j}-\frac{i}{\hbar}[\hat H_s+\frac{\De}{\w_o}\hat H_o+\hat H_r,\hat z_j]=\frac{i}{\hbar}[\hat H_{os},\hat z_j],\el{126}
\ba \dot{\hat p}_{j}-\frac{i}{\hbar}[\hat H_s+\frac{\De}{\w_o}\hat H_o+\hat H_r,\hat p_j]=\frac{i}{\hbar}[\hat H_{os},\hat p_j],\el{127}
where $j=s,o$. We define $C_{\hat p_j}=i[\hat H_{os},\hat p_j]/\hbar$ and $C_{\hat z_j}=i[\hat H_{os},\hat z_j]/\hbar$. The commutation relations $[\hat H_{os},\hat p_j]$ and $[\hat H_{os},\hat z_j]$ can not be determined without knowing the explicit form of the $\hat H_{os}$. However, we can guess that $C_{\hat z_j}$ and $C_{\hat p_j}$ are a function of $\hat z_s$, $\hat z_o$, $\hat p_s$, and $\hat p_o$ as $\hat H_{os}$ is also a function of the same variables.
\par As we are interested in measuring $\hat z_s$, without loss of generality, we assume that $\hat H_{os}$ is independent of $\hat p_s$. These assumptions not only simplify \eq{129} and \eq{128} but also represent the realistic experimental scenario in optical metrology. For example: the position (or $\hat z_s$) coupling comes naturally as any OSHO coupled to optical oscillator experiences position dependent electromagnetic field because of the wave nature of light~\cite{shepherd-90,abram}. Hence \eq{126} and \eq{127} can be rewritten as
\ba m_s(\ddot{\hat z}_s+\g_s\dot{\hat z}_s+\w_s^2\hat z_s)=C_{\hat p_s}(\hat z_o,\hat p_o,\hat z_s)+\hat s,\el{140}
\ba\dot{\hat p}_o+\g_o\hat p_o+m_o\w_o\De\hat z_o=C_{\hat p_o}(\hat z_o,\hat p_o,\hat z_s)+\hat o,\el{141}
where $\g_j$ is the damping rate and $\hat{j}$ is the corresponding noise operator~\cite{connell-88,vitali-01}. We adopt the following notation throughout this letter: Any arbitrary quantum operator $\hat A$ is written~\cite{zettili-text} as a sum of its expectation term $\bar A$ and its quantum fluctuation term $\hat\de_A$. Hence, by applying the Taylor expansion upto first order of quantum fluctuations, the dynamics of the quantum fluctuations in \eq{140}, and \eq{141} are given as
\ba\begin{split} m_s(\ddot{\hat\de}_{z_s}+\g_s\dot{\hat\de}_{z_s}+\w_s^2\hat\de_{z_s})=\um{j=s,o}{}\pd{C_{\hat p_s}}{\hat z_j}\big|_{\substack{\hat z_j=\bar z_j\\\hat p_j=\bar p_j}}\hat\de_{z_j}\\+\pd{C_{\hat p_s}}{\hat p_o}\big|_{\substack{\hat z_j=\bar z_j\\\hat p_j=\bar p_j}}\hat\de_{p_o}+\hat\de_s,\lbl{129}\end{split}\ea
\ba\begin{split}\dot{\hat\de}_{p_o}+\g_o\hat\de_{p_o}+m_o\w_o\De\hat\de_{z_o}=\um{j=s,o}{}\pd{C_{\hat p_o}}{\hat z_j}\big|_{\substack{\hat z_j=\bar z_j\\\hat p_j=\bar p_j}}\hat\de_{z_j}\\+\pd{C_{\hat p_o}}{\hat p_o}\big|_{\substack{\hat z_j=\bar z_j\\\hat p_j=\bar p_j}}\hat\de_{p_o}+\hat\de_o,\lbl{128}\end{split}\ea
Note the partial derivatives in \eq{129} and \eq{128} are a function of expectation terms only (no quantum terms). The transient behavior of the expectation terms is damped out on a time scale $t_e\gg1/\g_j$ leading to a steady state (any fast oscillation at optical frequencies is eliminated by the rotating wave approximation). Hence all the partial derivatives in \eq{129} and \eq{128} become time independent because of decoherence. As all the partial derivatives in \eq{129} and \eq{128} are a function of expectation terms only, from now onward, for notation simplicity, we shall drop indicating that the partial derivatives are evaluated at expectation terms. As a result, by using the Fourier transform definition $\mathfrak{F}(\hat A)=\int_{-\infty}^{\infty}\hat Ae^{i\w t}\dv t/\sqrt{2\pi}=\hat A(\w)$ with $\w$ as Fourier frequency, \eq{129} and \eq{128} imply that
\ba\hat\de_{z_s}(\w)=\pd{C_{\hat p_s}}{\hat z_o}\frac{\hat\de_{z_o}(\w)}{D_s(\w)}+\pd{C_{\hat p_s}}{\hat p_o}\frac{\hat\de_{p_o}(\w)}{D_s(\w)}+\frac{\hat\de_s(\w)}{D_s(\w)},\el{132}
\ba\begin{split}\hat\de_{p_o}(\w)(\g_o-i\w-\pd{C_{\hat p_o}}{\hat p_o})+m_o\W_o^2\hat\de_{z_o}(\w)=\hat\de_o(\w)\\+\hat\de_{z_{s}}(\w)\pd{C_{\hat p_o}}{\hat z_s},\lbl{133}\end{split}\ea
where $\W_o^2=\w_o\De-\pd{C_{\hat p_o}}{\hat z_o}\frac{1}{m_o}$, $D_s(\w)=m_s(\W_s^2-\w^2-i\g_s\w)$ with $\W_s^2=\w_s^2-\pd{C_{\hat p_s}}{\hat z_s}\frac{1}{m_s}$. The second terms on the left hand side (LHS) and the right hand side (RHS)of \eq{133} establish the presence of QBA. However, these terms arise from $\hat H_o$ and $\hat H_{os}$. Thus, unlike the oscillator in \eq{136}, the QBA in \eq{133} has contribution from the interaction between optical oscillator and OSHO as well. The optical oscillator can read the parameters of OSHO only if there is an interaction between them. This interaction generally involves exchange of forces which leads to perturbation of OSHO parameters. The quantum mechanical nature of the optical oscillator writes perturbations with quantum nature on to the OSHO and this manifests as QBA from the $\hat H_{os}$. We think the most popular example~\cite{caves-rp} for the QBA in optical metrology is the radiation pressure force noise in gravitational wave interferometer.
\par Coupled dynamics of optical oscillator and OSHO is given from \eq{132} and \eq{133} as
\begin{widetext}
\ba\hat\de_{p_o}(\w)=\frac{(\pd{C_{\hat p_o}}{\hat z_s}\pd{C_{\hat p_s}}{\hat z_o}\frac{1}{D_s(\w)}-m_o\W_o^2)\hat\de_{z_o}(\w)+\pd{C_{\hat p_o}}{\hat z_s}\frac{\hat\de_{s}(\w)}{D_s(\w)}+\hat\de_o(\w)}{(\g_o-i\w-\pd{C_{\hat p_o}}{\hat p_o}-\pd{C_{\hat p_o}}{\hat z_s}\pd{C_{\hat p_s}}{\hat p_o}\frac{1}{D_s(\w)})}.\el{134}\end{widetext}
On the RHS of \eq{134}, the first term in the numerator represents the QBA term. It reveals that QBA arises from previous measurement as well as from the interaction between the optical oscillator and the OSHO. The QBA in \eq{134} can be completely nullified if
\ba m_o\W_o^2=\pd{C_{\hat p_o}}{\hat z_s}\pd{C_{\hat p_s}}{\hat z_o}\frac{e^{i\tan^{-1}\E}}{\sqrt{m_s^2(\W_s^2-\w^2)^2+m_s^2\g_s^2\w^2}},\el{135}
where $\E=\g_s\w/(\W_s^2-\w^2)$. The LHS of \eq{135} comes from optical spring constant as well as $\hat H_{os}$ while the RHS comes from $\hat H_{os}$. Satisfying \eq{135} implies designing the measurement process such that the QBA perturbations from $\hat H_{os}$ are nullified by using the restoring force of the optical oscillator. The condition in \eq{135} can not be satisfied perfectly as the LHS is real and the RHS is complex, which means complete elimination of QBA is not possible using this method. Nevertheless, \eq{135} can lead to significant reduction in QBA if its real part is zero and its imaginary part is small (i.e., $\E\ll1$). For $\E\ll1$, the QBA in \eq{134} can be suppressed if
\ba m_o\W_o^2m_s\W_s^2=\pd{C_{\hat p_s}}{\hat z_o}\pd{C_{\hat p_o}}{\hat z_s}.\el{168}
Equation~(\ref{168}) is nothing but the real part of \eq{135} when $\E\ll1$. Using \eq{168} and \eq{134}, we can write 
\ba S_{p_op_o}\approx\frac{S_{z_oz_o}\big(\frac{\E}{m_s\W_s^2}\pd{C_{\hat p_s}}{\hat z_o}\pd{C_{\hat p_o}}{\hat z_s}\big)^2+S_{ss}\big(\frac{1}{m_s\W_s^2}\pd{C_{\hat p_o}}{\hat z_{s}}\big)^2+S_{oo}}{(\g_o-\pd{C_{\hat p_o}}{\hat p_o}-\frac{1}{m_s\W_s^2}\pd{C_{\hat p_s}}{\hat p_o}\pd{C_{\hat p_o}}{\hat z_s})^2},\label{167}\ea
where $\bra\hat\de_B(\w)\hat\de_B(\w')\ket=S_{BB}\de(\w+\w')$ with $B=p_o,z_o,s,o$. On the RHS of \eq{167}, in the numerator, the first, second, and third terms gives noise spectral densities from QBA, OSHO reservoir and optical oscillator reservoir, respectively. As $\E\to0$, QBA noise in \eq{167} goes to zero without affecting other terms. The condition $\E\ll1$ can be realized experimentally for $\W_s\gg\w$ and $\W_s\gg\g_s$. As $S_{z_oz_o}$, $S_{p_op_o}$, and all the partial derivatives in \eq{167} are independent of $\w$, the QBA noise in \eq{167} goes to zero as $\E\to 0$. Thus $\E\ll1$ is a necessary condition which limits QBA suppression using \eq{168} only to the low frequency regime. We neglected the cross-correlation terms in \eq{167} by setting $\left\bra\hat\de_{z_o}\hat\de_s\right\ket=\left\bra\hat\de_{z_o}\hat\de_o\right\ket=0$. Even if these cross-correlations are not neglected, noise contribution from them becomes zero upon frequency symmetrization. Despite of the complex nature of \eq{135}, \eq{168} gives an experimentally feasible criteria to extract information from $\hat H_s$ without QBA in the low frequency regime by measuring $\hat p_o$ of the optical oscillator. The advantage of using an optical oscillator becomes obvious once we recognize that $\hat p_o$ is proportional to the optical oscillator's phase quadrature which can be measured in a homodyne setup~\cite{grangier}. Using an optical oscillator as a meter not only allows us to measure $\hat p_o$ in a relatively simple way but also to coherently couple with variety of OSHOs~\cite{ashkin,gil-santos,shkarin,esslinger} like a mechanical oscillator~\cite{kippenberg-om} or atoms in harmonic trap~\cite{schreppler} etc. Hence the \eq{134}, \eq{168} and \eq{167} can be applied to many systems for QBA evasion in low frequency regime.
\par Before we go any further, it is useful to note that the condition in \eq{135} is derived by assuming that $\hat p_s$ is not coupled to the OSHO. This is a reasonable but not a necessary assumption in the context of OSHO or quantum optical metrology. Thus it is natural to wonder about QBA evasion when $\hat p_s$ is coupled to the optical oscillator. Equation-\ref{129} and \eq{128} is general enough to include both $\hat{z}_s$ and $\hat p_s$ coupling to the optical oscillator. Hence starting from \eq{129}, 
by following the same steps that led to \eq{135}, we obtain that the condition for QBA evasion as
\ba m_o\W_o^2=\frac{\pd{C_{\hat z_s}}{\hat z_o}\pd{C_{\hat p_o}}{\hat p_{s}}}{\frac{\w^2}{m_s\w_s^2}-\left(\frac{1}{m_s}+\pd{C_{\hat z_s}}{\hat p_s}\right)+i\frac{\g_s\w}{m_s\w_s^2}}\el{142} 
when $\hat p_s$, instead of $\hat z_s$, is coupled to the optical oscillator. Derivation \eq{142} is given in the supplementary material. Again by noting that, in \eq{142}, the LHS is real and the RHS is complex, QBA evasion can be achieved in the low frequency regime, i.e., $\w\to0$ when
\ba m_o\W_o^2\left(\frac{1}{m_s}+\pd{C_{\hat z_s}}{\hat p_s}\right)=-\pd{C_{\hat z_s}}{\hat z_o}\pd{C_{\hat p_o}}{\hat p_{s}}.\el{137}
\par Using \eq{135} and \eq{134}, as a direct application of the theory developed so far, we will derive the conditions for achieving radiation pressure force noise suppression in optomechanics~\cite{kippenberg-om}.
\begin{figure}[htb]\centering\includegraphics{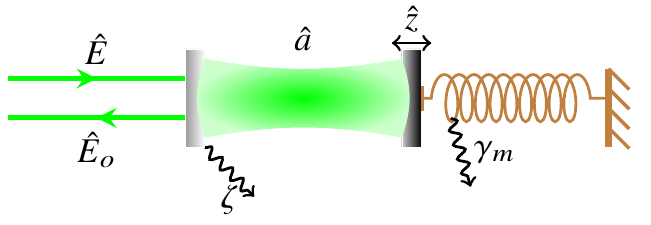}\caption{A generic Optomechanical cavity. The cavity field $\hat a$, when reflected from the optomechanical oscillator, displaces the position $\hat z$ of the optomechanical oscillator. The change in the length of the optical cavity modifies the optical response of the optomechanical cavity. $\g_m$ and $\zeta$ are the decay rates of optomechanical mirror and optomechanical cavity, respectively.}\label{f13}\end{figure}
\par Figure-\ref{f13} represents a generic optomechanical system. The optomechanical Hamiltonian $\hat H_{om}$ is given~\cite{law} as
\ba\hat H_{om}=\frac{\hat p^2}{2m}+\frac{m\w_m^2\hat z^2}{2}+\left(\frac{\hat P^2}{2M}+\frac{M\w_a^2\hat Z^2}{2}\right)\left(\frac{\De_a}{\w_a}-\frac{\hat z}{l}\right),\el{169}
where $\hat p$, $\hat z$, $m$, and $\w_m$ are the momentum, position, mass and eigen frequency of the mechanical oscillator, respectively. $\hat P=i\hbar\w_a(\hat a^\+-\hat a)/\sqrt 2c$, $\hat Z=c(\hat a^\++\hat a)/\sqrt 2\w_a$ and $M=\hbar\w_a/c^2$ with $\w_a$, $\hat a$, $\hat a^\+$, as the Eigen frequency, annihilation operator and creation operator for the cavity mode, respectively. $\De_a$ is the detuning between drive and cavity Eigen frequency, $c$ is the speed of light in vacuum and $[\hat Z,\hat P]=i\hbar$. The first two terms on the RHS of \eq{169} represent Hamiltonian of the mechanical oscillator, which plays the role of system oscillator, from which information is extracted by the cavity field which plays the role of optical oscillator. The third term on the RHS of \eq{169} represents the Hamiltonian for the cavity field and its interaction with the mechanical oscillator. We didn't include the reservoir terms in \eq{169}. By noting that $\hat p$, $\hat z$, $\hat P$, and $\hat Z$ are similar to $\hat p_s$, $\hat z_s$, $\hat p_o$, and $\hat z_o$, respectively, we can immediately apply \eq{135} to derive the condition for achieving QBA evasion in optomechanics. By realizing that $C_{\hat p_o}$ is similar to $C_{\hat P}=M\w_a^2\hat Z\hat z/l$, and $C_{\hat p_s}$ is similar to $C_{\hat p}=(\hat P^2/2M+M\w_a^2\hat Z^2/2)/l$, we can evaluate the condition in \eq{135} for $\hat H_{om}$ as
 \ba\frac{\hbar\w_a^2(\bar a+\bar a^*)^2}{2l^2D(\w)}=\left(\De_a-\w_a\frac{\bar z}{l}\right).\el{122} 
Here $D(\w)=m(\w_m^2-\w^2-i\g_m\w)$ with $\g_m$ as the optomechanical oscillator damping, $\bar a^*$ is the complex conjugate of $\bar a$, and $\bar Z=(\bar a+\bar a^*)\sqrt{\hbar/2M\w_a}$. It can be directly verified whether \eq{122} can suppress the optomechanical radiation pressure force noise by cross-checking with the dynamical equations~\cite{aspelmeyer-rmp,sankar-njp2} of $\hat a$. A discussion on more details about practical implementation of \eq{135} on $\hat H_{om}$ will be interesting, but that is beyond the scope of this letter and the interested readers may find such details in \cite{sankar-ol2}. Application of \eq{135} to \eq{169} is only one example. In fact, we can also apply \eq{135} to non-linear optomechanics where $\hat z$ is kept to all orders. The Hamiltonian $\hat H_{n}$ for the non-linear optomechanics is given as
\ba\hat H_{n}=\frac{\hat p^2}{2m}+\frac{m\w_m^2\hat z^2}{2}+\Big(\frac{\hat P^2}{2M}+\frac{M\w_a^2\hat Z^2}{2}\Big)\Big(1+\frac{\hat z}{l}\Big)^{-1}.\el{162}
Now by applying \eq{135} to \eq{162}, one can derive the condition for avoiding QBA in non-linear optomechanics too.
\par In this paragraph we compare our techniques with other prominent methods like QND, squeezing, Coherent quantum noise cancellation (CQNC), quantum mechanics free subsystems (QMFS), and variational measurements. The sufficient condition~\cite{braginsky-80,braginsky-rmp,grangier} for QND measurement is that the measured variable has to commute with the total Hamiltonian. While this can eliminate QBA completely, it is very rare to find systems for which the Hamiltonian commute with the measured variable. Sending squeezed vacuum through the empty port of the interferometer is another technique~\cite{schnabel-17,purdy-20,sankar-njp1} to suppress the QBA, if the squeeze angle is appropriately optimized. The strength of squeeze parameter determines the effectiveness of this method. To our knowledge, highest squeezing achieved so far is $15$\,db~\cite{schnabel-16}. Achieving high quality squeezing is the main challenge in this technique. The CQNC~\cite{caves-12,Likai-1} uses an auxiliary system which is coupled to the main system. The auxiliary system is to be synthesized~\cite{bariani,motazedifard-16,naderi-22} such that the QBA noise from the main system cancel with the noise from the auxiliary system. The success of CQNC depends on the finetuning of the auxiliary system. Another method  to avoid QBA is through QMFS~\cite{moller,sillanpaa-21}. As the QBA arises because of the interplay between the canonically conjugate variables, an effective negative mass system is created to satisfy the relation $[\hat x_1+\hat x_2,\hat p_1-\hat p_2]=0$. Where $\hat x_1$, $\hat x_2$ and $\hat p_1$, $\hat p_2$ are positions and momentum of two systems, respectively. The negative mass leads to minus sign before $\hat p_2$. In contrast to all these methods, a QBNM works by choosing a meter with intrinsic restoring force. The restoring force of the optical oscillator is used to reduce the randomness coming from the QBA. It is also possible to combine QBNM with other established techniques like squeezing etc to improve the overall effectiveness of both the methods.
\par Theory for quantum back-action nullifying meter is developed. By assuming the meter as an optical oscillator, a new method to achieve QBA evasion in the low frequency regime in quantum optical metrology is developed. Evasion of QBA is achieved by using the restoring force of the optical oscillator to counter the perturbation induced by the QBA force. Application of QBNM for nullifying the quantum radiation pressure force noise in optomechanics is presented.
\section{Funding}
This work is supported by Science and Engineering Research Board, India, under the grant no: SRG/2020/001167. It is also supported by the National Natural Science Foundation of China (Grants No. 12074030, and No. U1930402).
\bibliography{references}
\end{document}